\documentclass[times, twoside]{zHenriquesLab-StyleBioRxiv}
\usepackage{blindtext}
\usepackage{IEEEtrantools}

\leadauthor{Manzaneque} 

\begin{document}

\title{Resolution Limits of Resonant Sensors with Duffing Non-Linearity}
\shorttitle{Resolution Limits of Resonant Sensors with Duffing Non-Linearity}

\author[ab]{Tomás Manzaneque}
\author[b]{Murali K. Ghatkesar}
\author[b]{Farbod Alijani}
\author[bc]{Minxing Xu}
\author[bc]{Richard A. Norte}
\author[bc]{Peter G. Steeneken}

\affil[a]{Department of Microelectronics, Delft University of Technology}

\affil[b]{Department of Precision and Microsystems Engineering, Delft University of Technology}

\affil[c]{Department of Quantum Nanoscience, Delft University of Technology}

\maketitle

\begin{abstract}
The resolution of resonant sensors is fundamentally limited by the presence of noise. Thermomechanical noise, intrinsic to the resonator, sets the ultimate sensor performance when all other noise sources have been eliminated. For linear resonators, the sensing resolution can always be further improved by increasing the driving power. However, this trend cannot continue indefinitely, since at sufficiently high driving powers non-linear effects emerge and influence the noise performance. As a consequence, the resonator's non-linear characteristics play an inextricable role in determining its ultimate resolution limits. Recently, several works have studied the characteristic performance of non-linear resonators as sensors, with the counter intuitive conclusion that increasing the quality factor of a resonator does not improve its sensing resolution at the thermomechanical limit. In this work we further analyze the ultimate resolution limits, and describe different regimes of performance at integration times below and above the resonator's decay time. We provide an analytical model to elucidate the effects of Duffing non-linearity on the resolution of closed-loop sensors, and validate it using numerical simulations. In contrast to previous works, our model and simulations show that under certain conditions the ultimate sensing resolution of a Duffing resonator can be improved by maximizing its quality factor. With measurements on a nanomechanical membrane resonator, we experimentally verify the model and demonstrate that frequency resolutions can be achieved that surpass the previously known limits.
\end{abstract}


\begin{corrauthor}
t.manzanequegarcia\at tudelft.nl
\end{corrauthor}

\section*{Introduction}
The use of mechanical resonators as precise sensors for mass, force and other physical parameters has been fueled by recent advances in miniaturization. This is because a lower resonator mass leads to a higher responsivity, i.e. a larger shift of the resonance frequency for a given stimulus \cite{ekinci2004ultimate}. However, the detection limit of resonant sensors is not only determined by the responsivity, but also by the frequency resolution, which is defined as the smallest resonance frequency change that can be detected. In order to be resolved, a frequency change needs to be larger than the stochastic variations observed when measuring the resonance frequency, which are commonly characterized by the Allan deviation $\sigma_y$ \cite{allan1966statistics,barnes1971characterization}. The Allan deviation of a resonant sensor results from the noise sources inherent to the resonator, its environmental conditions, and the noise from the readout \cite{ekinci2004ultimate,cleland2002noise,sansa2016frequency,manzaneque2020method}. When all environmental and readout noise sources are eliminated, the fundamental resolution limit at finite temperatures is determined by the resonator's thermomechanical noise. The corresponding Allan deviation, which depends on the integration time of the measurement, has been well established for mechanical resonators in the linear regime \cite{ekinci2004ultimate,demir2019fundamental}.

\begin{figure}
\centering
\includegraphics[width=1\linewidth]{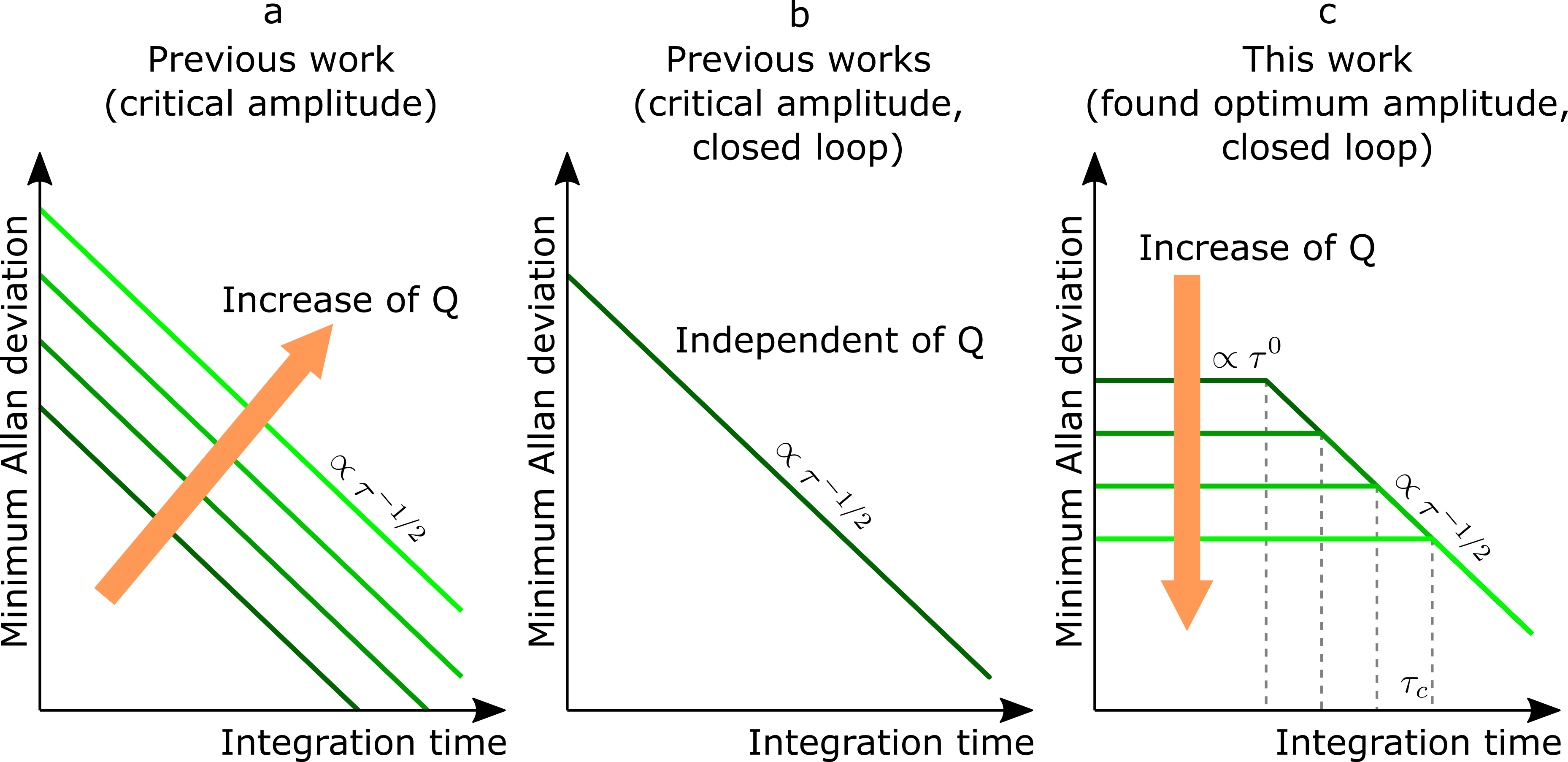}
\caption{Log-log plots of the minimum Allan deviation that can be obtained by a thermomechanically-limited Duffing resonator when optimizing the actuation power, as a function of the integration time $\tau$. Each curve corresponds to a different value of the quality factor. (a) Limit found by Roy et al.~\cite{roy2018improving} by assuming the minimum Allan deviation is reached for actuation at the critical amplitude. (b) Limit found by Olcum et al.~\cite{olcum2014weighing} for high quality factor resonators, and by Demir and Hanay~\cite{demir2019fundamental} for any quality factor but assuming closed-loop operation. Both works use the assumption that the minimum Allan deviation is reached at or close to the critical amplitude. (c) Limit for Duffing resonators under closed-loop operation, by using the optimum actuation amplitude found in this work, different for each integration time. For short integration times, the minimum Allan deviation is independent of the integration time. For long integration times, the minimum Allan deviation shows a $\tau^{-1/2}$ dependence and matches the result in (b). The transition between both regimes occurs at $\tau_c=\sqrt{3} Q/(\pi f_0$).}
\label{fig:intro}
\end{figure}

The linear harmonic oscillator model does not predict a lower limit for the Allan deviation, since increasing the actuation power, and so the signal to noise ratio, always results in a lower Allan deviation $\sigma_y$ for a given integration time. Several authors have pointed out that this trend cannot continue indefinitely, due to the non-linear effects in the resonator response at high enough actuation powers \cite{ekinci2004ultimate, roy2018improving, demir2019fundamental, molina2021high,olcum2014weighing}. For mechanical resonators at the micro- and nano-scale, non-linear effects usually manifest as a cubic stiffness in the resonator dynamics, commonly known as the Duffing non-linearity. This characteristic comes with conversion of amplitude noise into phase noise that enhances with the actuation power, worsening the frequency resolution. Therefore, it is expected that Duffing resonators present an optimum power level as a result of the trade-off between signal to noise ratio and amplitude-phase noise conversion. Following this rationale, it has been assumed that the minimum Allan deviation is obtained for a characteristic oscillation amplitude at which the non-linear effects become dominant, known as the critical amplitude or onset of non-linearity\cite{roy2018improving,demir2019fundamental,olcum2014weighing}. Using the fact that this critical amplitude is inversely proportional to the quality factor $Q$, it has been proposed that the minimum Allan deviation is independent of $Q$ \cite{demir2019fundamental,olcum2014weighing}, see Fig.~\ref{fig:intro}. Using the same argument, Roy et al. \cite{roy2018improving} concluded that the minimum Allan deviation can be even reduced by decreasing the quality factor of a resonator, see Fig.~\ref{fig:intro}(a). Differently from the previous ones, this study did not include the dynamics of the closed-loop controller often used to drive resonant sensors \cite{demir2019fundamental}. With or without the closed-loop dynamics, the results cited noticeably contrast with  historical efforts of the resonant sensor community to increase the quality factor of mechanical resonators to improve resolution.


In this work, we experimentally and theoretically analyze the ultimate frequency resolution of Duffing resonators in closed-loop operation. Our experimental results show that sensor resolution can be improved beyond the previously determined limits \cite{demir2019fundamental,olcum2014weighing} for integration times $\tau$ shorter than a critical time constant $\approx Q/f_0$, where $Q$ is the quality factor and $f_0$ is the linear resonance frequency. To gain insight into the problem, the effect of white noise on the Allan deviation of a Duffing resonator in closed loop is modelled using perturbation theory. The developed model, based on linearization of the amplitude-phase space of the Duffing resonator, closely reproduces our experimental findings. In addition, the theory provides useful expressions to determine the minimum Allan deviation limit in the non-linear regime, and the optimal actuation level needed to reach it, which turns out to depend on the integration time.

Our results show that two regimes can be identified in which the minimum Allan deviation of a Duffing resonator exhibits distinct behaviours. For fast sensing (short integration times), the ultimate resolution limit is independent of the integration time, and can be improved by increasing the resonator's quality factor. For slow sensing (long integration times), the ultimate resolution limit does not depend on the quality factor as long as all other resonator characteristics are left unchanged. Fig. \ref{fig:intro}(c) shows a graphical representation of the determined resolution limits and their dependence on the quality factor and the integration time.

In the following, we first present an experimental study of the frequency resolution of a non-linear silicon nitride nanomechanical membrane and its dependence on driving power and integration time. Then we theoretically analyze the resolution limits of Duffing resonators and derive analytic expressions for the minimum Allan deviation, for limiting cases of short and long integration times. Finally, we perform numerical simulations of the Duffing resonator in closed loop to validate the analytical model, and confirm the conclusions without the linearization approximations introduced in the derivations.


\section*{Experimental Results}
\begin{figure*}
\centering
\includegraphics[width=\textwidth]{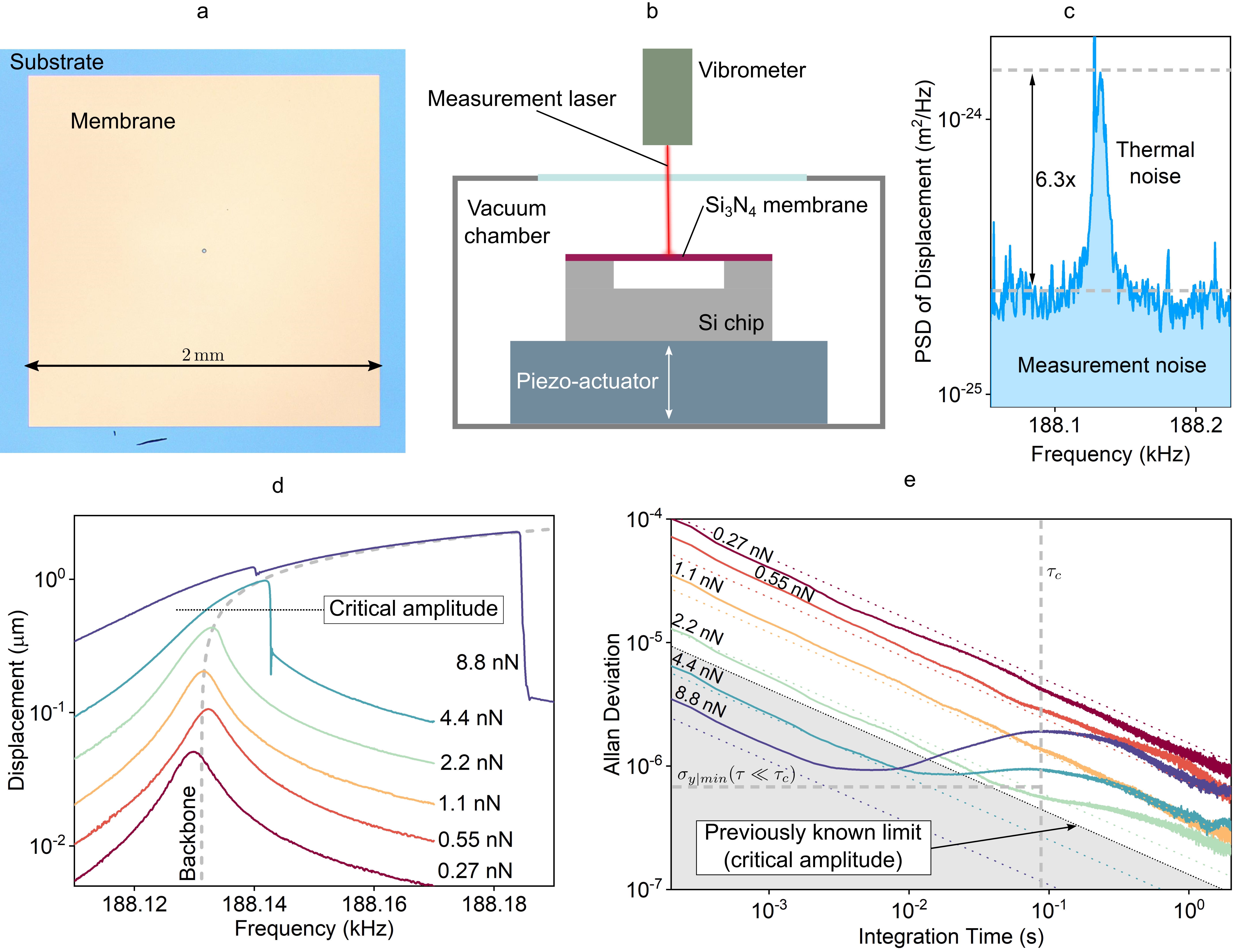}
\caption{(a) Top-view optical picture of the \ce{Si3N4} membrane used in the experiments. (b) Cross-section representation of the chip containing the membrane in the experimental setup. (c) Measured noise spectrum in absence of actuation. The peak corresponds to the thermomechanical noise of the membrane's fundamental resonance. By fitting to a Lorentzian, the resonance frequency, quality factor, modal stiffness and modal mass gathered in Table~\ref{tab:parameters} were calculated. (d) Frequency sweeps around the fundamental resonance for different actuation levels. For low levels, the membrane behaves as a linear resonator. For \SI{2.2}{\nano\newton} and above, a stiffening effect is observed that produces bifurcation for \SI{4.4}{\nano\newton} and above. The maxima of the different curves were fitted to the backbone curve of Duffing resonators, from which the Duffing parameter $\gamma$ in Table~\ref{tab:parameters} was obtained. The critical amplitude calculated from $\gamma$ and $Q$ \cite{febbo2013critical}, $a_{\rm crit}=\SI{0.59}{\micro\metre}$, is indicated. (e) Allan deviation for the same actuation levels as in (d), when the membrane is driven by a PLL controller at the displacement maxima. The solid lines represent measurements, where white force noise with PSD $S_{\rm 0}= \SI{2.5e-21}{\square\newton\per\hertz}$ was added to the actuation to ensure the response is dominated by white force noise. This force noise resembles the thermomechanical noise the resonator would experience at temperatures higher than room temperature. The dotted colored lines represent the theoretical values for the membrane assuming linear response under the experimental conditions. The dotted black line indicates the previously known limit for the Allan deviation, defined by actuation at the critical amplitude. The horizontal dashed line indicates the lower boundary for the Allan deviation found by our model at integration times below $\tau_c$.}
\label{fig:results}
\end{figure*}

A series of experiments was performed on a square silicon nitride membrane with a thickness of $\SI{92}{\nano\metre}$ and an area of \SI[product-units=power]{2 x 2}{\milli\metre\squared} that was suspended on a silicon chip, see Figs.~\ref{fig:results}(a,b). The chip was mounted on a piezo-actuator inside a vacuum chamber at a pressure of \SI{2}{\milli\pascal} (see Methods). The motion in the center of the membrane was recorded by Doppler laser vibrometry. The measurement setup was sensitive enough to characterize the resonator's thermomechanical displacement noise in the absence of piezo-actuation, whose power spectral density (PSD) spectrum is shown in Fig.~\ref{fig:results}(c). The figure shows a peak at 188 kHz that corresponds to the fundamental vibration mode of the membrane. The peak, with an amplitude $6.3$ times larger than the measurement noise floor allowed to extract the linewidth. The parameters defining the linear resonance, listed in Table~\ref{tab:parameters}, were obtained by fitting the PSD data to a harmonic oscillator model \cite{hauer2013general}. Then, upward frequency sweeps were performed around the found resonance with different actuation forces, with the results plotted in Fig.~\ref{fig:results}(d). As the actuation force increases, a hardening effect becomes evident from the bending of the resonance peak to the right. For an actuation force of \SI{4.4}{\nano\newton} and above, a sudden jump in the displacement amplitude is observed, which indicates that the critical amplitude has been surpassed, i.e. the resonator shows bifurcation \cite{febbo2013critical}. The displacement maxima with amplitude $a_p$ for each actuation level were fitted to the backbone curve of Duffing resonators \cite{nayfeh2008nonlinear},
\begin{equation}\label{eq:backbone}
    \omega_p=\omega_0 \left( 1+\frac{3}{8}\gamma a_p^2  \right),
\end{equation}
with $\omega_p$ the angular frequency at which amplitude $a_p$ is reached, and $\omega_0$ the angular resonance frequency in the linear regime. The obtained curve, plotted in Fig.~\ref{fig:results}(d), allowed to extract the value of the Duffing coefficient $\gamma$ shown in Table \ref{tab:parameters}, see also \eqref{eq:duffing_res_2}.

Next, to evaluate the effects of thermomechanical noise on the Allan deviation of the closed-loop-driven Duffing resonator, experiments in the presence of white force noise were performed. For that, a phase locked-loop (PLL) control scheme was set to continuously drive the resonator at resonance (see Methods), with PLL parameters set using the method from \cite{olcum2015high}. The chosen configuration ensures that the observed Allan deviation is independent of the PLL parameters for integration times longer than \SI{100}{\micro\second}. The controller was set to target a phase shift of $-\pi/2$ in the resonator's response, where phase shifts introduced by other components were compensated for. This condition ensures that the device is driven at $\omega_0$ when the applied force is in the linear regime, and at the non-linear resonance frequency $\omega_p$ when the resonator enters the non-linear regime \cite{denis2018identification}.

The experimental analysis of resonators at their thermomechanical limits is hindered by resonance frequency fluctuations \cite{sansa2016frequency} and the presence of measurement noise introduced by instrumentation. To illustrate the challenge to study the Allan deviation of micro- and nano-mechanical resonators in the absence of measurement noise, we note that to reach that condition the thermomechanical noise peak in Fig.~\ref{fig:results}(c) would have to exceed the background PSD at $f<f_0$ by a factor $Q^2=\SI{9e8}{}$ \cite{miller2018effective}, which is much higher than the experimental ratio of 6.3. To overcome this problem, we ensure that the resonator operates in the regime where its response is dominated by white force noise by using the piezo to add a random force, with white spectral distribution and PSD $S_0$, to the sinusoidal actuation force provided by the PLL system. According to the fluctuation-dissipation theorem, the thermomechanical noise of a system is fully accounted by a force of such type and power spectral density (one-sided, on a per hertz basis) \cite{callen1951irreversibility}
\begin{equation}\label{eq:S_thm}
    S_{\rm thm}=4 c k_B T,
\end{equation}
where $k_B$ is the Boltzmann constant, $T$ is the absolute temperature and $c$ is the dissipation constant of the system. For either linear or Duffing resonators, $c=\sqrt{k_1 m}/Q$, where $k_1$ is the linear modal stiffness, $m$ is the modal mass, and $Q$ is the quality factor. By adding white noise with PSD $S_0$, we emulate the thermomechanical noise the resonator would experience at a higher effective temperature $T_{\rm eff}= S_0/(4c k_B)$. Making $S_0$ much larger than the other noise sources ensures that the Allan deviation is limited by this artificial thermomechanical noise. Thus, our experiments capture the ultimate situation where thermomechanical noise dominates the frequency resolution, such that contributions of instrumentation and fluctuations on the operation conditions can be neglected. In all of the Allan deviation experiments, the amplitude of the white noise was set to a constant value of $S_{\rm 0}= \SI{2.5e-21}{\square\newton\per\hertz}$.

\begin{table*}[tbhp]
    \centering
    \caption{Resonator parameters obtained from experiments. The resonance frequency, quality factor, modal stiffness and modal mass are obtained from the data in Fig.~\ref{fig:results}(c). The Duffing coefficient is obtained from the backbone curve in Fig.~\ref{fig:results}(d).}
    \begin{tabular*}{\textwidth}{c @{\extracolsep{\fill}} cccc} \hline    
    Resonance frequency&Quality factor&Modal stiffness&Modal mass& Duffing coefficient\\ 
    $f_0$&$Q$&$k_1$&$m$&$\gamma$\\
    \SI{188.1}{\kilo\hertz}&\SI{30e3}&\SI{320}{\newton\per\metre}&\SI{229}{\nano\gram}&\SI{1.47e8}{\per\square\metre}\\
    \hline
    \end{tabular*}
    \label{tab:parameters}
\end{table*}

In  Fig.~\ref{fig:results}(e) (solid lines), we report the Allan deviations measured at the same actuation forces as in Fig.~\ref{fig:results}(d). The diagonal dotted lines represent the calculated Allan deviation for linear resonators in closed-loop as follows~\cite{demir2019fundamental}:
\begin{equation}\label{eq:sigma_linear}
    \sigma_{y0}(\tau)=\frac{1}{a_x} \sqrt{\frac{k_B T}{ m Q \omega_0^3  \tau }},
\end{equation}
where $\tau$ is the integration time and $a_x$ is the displacement amplitude of the oscillation. Note that the phase condition $\phi=-\pi/2$ set in the experiments ensures the driving frequency equals $\omega_p$ and therefore $a_x$ equals the peak amplitude $a_p$. It is observed that, for force amplitudes in the linear range of operation ($\SI{0.27}{\nano\newton}$, $\SI{0.55}{\nano\newton}$ and $\SI{1.1}{\nano\newton}$), the measured Allan deviation shows good agreement with \eqref{eq:sigma_linear}.  
At actuation force levels of $\SI{2.2}{\nano\newton}$ and above, where the resonator response in Fig.~\ref{fig:results}(d) is seen to enter the non-linear regime, the measured Allan deviation in Fig.~\ref{fig:results}(e) is seen to differ strongly in behaviour from the linear model, with three distinguishable regimes. In the first regime, for short integration times $\tau$, the Allan deviation behaves as predicted by the linear model, with a $\tau^{-1/2}$ dependence like in \eqref{eq:sigma_linear} (indicated by dotted lines in the figure). Note that the Allan deviation is not affected by the Duffing non-linearity in this regime, even for actuation forces that exceed the critical amplitude (\SI{4.4}{\nano\newton} and \SI{8.8}{\nano\newton}). At intermediate values of $\tau$, the Allan deviation increases significantly. Finally, in the third regime at integration times longer than $\approx Q/f_0$, the Allan deviation reduces again with a $\tau^{-1/2}$ trend, although at significantly higher levels than obtained from \eqref{eq:sigma_linear}. In this last regime, the data corresponds well to the theoretical prediction from~\cite{demir2019fundamental}, with the lowest Allan deviation being found at a driving force of $\SI{2.2}{\nano\newton}$, which is the force that is closest to the critical amplitude in Fig.~\ref{fig:results}(d). However, this driving force does not give the lowest Allan deviation for all integration times, since the curves at \SI{4.4}{\nano\newton} and \SI{8.8}{\nano\newton} show lower values in the region $\tau<Q/f_0$. In addition, these curves show local minima at an approximately constant value of the Allan deviation. In the next sections, this behaviour of the Allan deviation in the non-linear regime is analyzed further by theory and simulations.

\section*{Theory and Model}
For a driven Duffing resonator the modal displacement $x$ and modal force $g$ are related by the second-order differential equation
\begin{equation}\label{eq:duffing_res}
    m\ddot{x}(t)+c\dot{x}(t)+k_1 x(t) +k_3 x^3(t) =g(t),
\end{equation}
where $t$ represents the time and $k_3$ is the cubic stiffness of the vibration mode. This equation can be simplified by normalizing the time, $\hat{t}=t\omega_0$, and the force, $\hat{g}=g/k_1$, resulting in
\begin{equation}\label{eq:duffing_res_2}
    \ddot{x}(\hat{t})+2\Gamma\dot{x}(\hat{t})+ x(\hat{t}) + \gamma x^3(\hat{t}) =\hat{g}(\hat{t}),
\end{equation}
where $\Gamma=1/(2Q)$ is the damping ratio and $\gamma=k_3/k_1$ is the Duffing coefficient.

\begin{figure*}
\centering
\includegraphics[width=\textwidth]{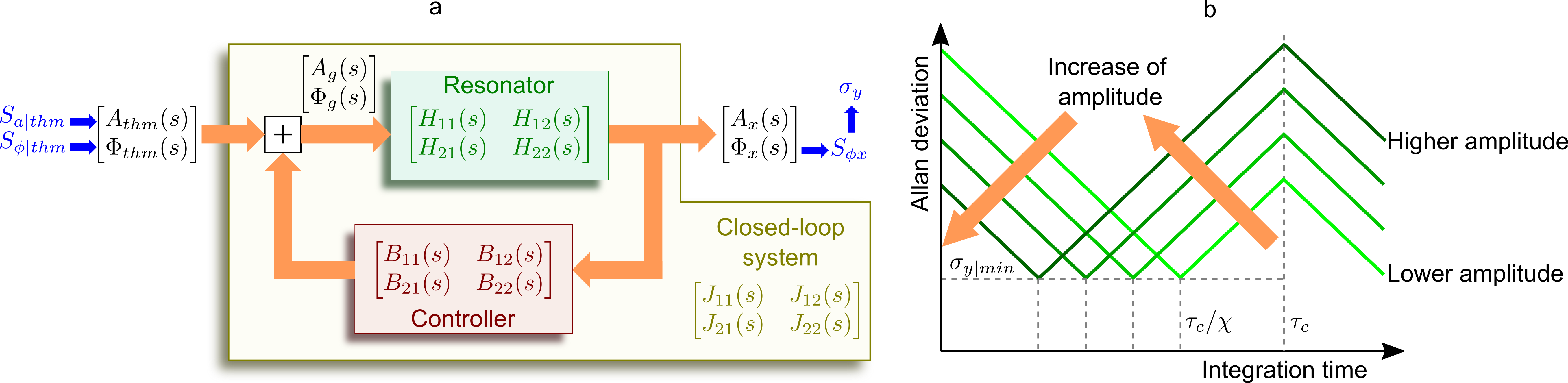}
\caption{(a) Laplace-domain block diagram of the linearized small-signal model of the resonator and the resonance-tracking control loop. The closed-loop system matrix $J(s)$ governs the conversion of amplitude and phase perturbations at the input (force) into amplitude and phase perturbations at the output (displacement). In blue, input and output magnitudes used in the analysis. Amplitude and phase random perturbations due to thermomechanical noise, with PSDs $S_{a|\textrm{thm}}$ and $S_{\phi|\textrm{thm}}$ respectively, translate into phase perturbations at the output with PSD $S_{\phi x}$, from which the Allan deviation ($\sigma_y$) is calculated. (b) Log-log plot of the Allan deviation predicted by the model in (a), for a Duffing resonator at its thermomechanical limit oscillating at $\phi=-\pi/2$. Each curve represents a different amplitude of oscillation. A characteristic minimum $\sigma_{y\textrm{|min}}$ is reached at an integration time $\tau_c/\chi$ that depends on the amplitude of oscillation. For $\tau\ll\tau_c/\chi$, the Allan deviation is governed by the linear response and the Duffing term can be neglected ($\gamma=0$).}
\label{fig:model}
\end{figure*}

Let the displacement and force be approximately harmonic functions of $\hat{t}$, $x(\hat{t})=a_x(\hat{t}) \sin{(\hat{\omega} \hat{t} + \phi_x(\hat{t}))}$ and $\hat{g}(\hat{t})=a_g(\hat{t})\sin{(\hat{\omega} \hat{t} +\phi_g(\hat{t}))}$ respectively, with $\hat{\omega}=\omega/\omega_0$. The amplitudes and phases are assumed to vary slowly, at characteristic times much shorter than the inverse of the resonance frequency $\omega_0$. In the high $Q$ limit, the slow dynamics of the resonator can be modelled by applying the method of averaging \cite{nayfeh2008nonlinear,kenig2012optimal}:
\begin{IEEEeqnarray}{CCLL}
\label{eq:slowdyn1}
    \dot{a}_x&=&-\Gamma a_x - \frac{1}{2} a_g \sin{\phi} &\equiv \eta_a\\
    \label{eq:slowdyn2}
    \dot{\phi}_x&=&\frac{3}{8} \gamma a_x^2 - \psi -\frac{1}{2} \frac{a_g}{a_x}\cos{\phi} &\equiv \eta_{\phi}\\
    \phi&=&\phi_x-\phi_g\label{eq:slowdyn3}
\end{IEEEeqnarray}
$\psi=\hat{\omega} -1$ is the detuning of the normalized actuation frequency,  where $\psi=0$ represents the tuned condition in which the resonator is actuated at $\omega_0$.

Equations (\ref{eq:slowdyn1}-\ref{eq:slowdyn3}) describe a system of non-linear differential equations that relate the amplitude and phase of the displacement to the amplitude and phase of the applied force. As detailed in Supplementary Note \ref{note:linearization}, this system can be linearized around a fixed point and translated into the Laplace domain to obtain a matrix of transfer functions $H(s)$ that relates the inputs to the outputs:
\begin{equation}\label{eq:laplace}
    \begin{bmatrix} A_x(s) \\ \Phi_x(s) \end{bmatrix}=
    \begin{bmatrix} H_{11}(s)&H_{12}(s) \\ H_{21}(s)&H_{22}(s) \end{bmatrix}
    \begin{bmatrix} A_g(s) \\ \Phi_g(s) \end{bmatrix}
\end{equation}
$A_x$, $\Phi_x$, $A_g$ and $\Phi_g$ represent the Laplace transforms of the small perturbations of $a_x$, $\phi_x$, $a_g$ and $\phi_g$ respectively around the fixed point, while $s$ denotes the Laplace variable. The matrix $H(s)$ depends on $\Gamma$, $\gamma$ and the chosen fixed point. 

In a closed-loop system, the resonator's phase shift and amplitude of oscillation can be set at arbitrary values. That is to say, we have control over $\phi$ and either $a_x$ or $a_g$ to define the fixed point. The variables of a particular fixed point, at which $\dot{a}_x=0$ and $\dot{\phi}_x=0$, are found from Eqs.~(\ref{eq:slowdyn1}-\ref{eq:slowdyn3}) to be related through
\begin{IEEEeqnarray}{CCL}
    \label{eq:fixed_point_equations_1}
    a_g&=&-\frac{2\Gamma a_x}{\sin\phi}\\
    \label{eq:fixed_point_equations_2}
    \psi&=&\frac{3}{8}\gamma a_x^{2}+\frac{\Gamma}{\tan \phi}.
\end{IEEEeqnarray}
In addition to defining the fixed point of the resonator model, the control loop influences the time evolution of the noise-induced amplitude and phase perturbations around the fixed point. Figure \ref{fig:model}(a) shows a block diagram in the Laplace domain of the closed-loop model for the perturbations. A closed-loop transfer function matrix $J(s)$ is defined from the resonator's open-loop transfer function matrix $H(s)$ and the transfer function matrix $B(s)$ of the feedback controller. The derivation of these matrices, that can be found in Supplementary Note \ref{note:linearization}, results in 
\begin{IEEEeqnarray}{CCLC}
\label{eq:cl1}
    J_{11}(s)&=&-\frac{\sin \phi}{2 (s+\Gamma)}& \;\;\;\;\;\;
    \\
\label{eq:cl2}
    J_{12}(s)&=&-\frac{\Gamma a_x }{\tan\phi(s+\Gamma)}&\;\;\;\;\;\;
    \\
\label{eq:cl3}
    J_{21}(s)&=&-\frac{3 \gamma a_x^{2} \sin \phi +4 s \cos \phi}{8 a_x s ( s + \Gamma)}&\;\;\;\;\;\;
    \\
\label{eq:cl4}
    J_{22}(s)&=&\frac{\Gamma \left(4 s \tan^2{\phi} -3 \gamma a_x^{2} \tan{\phi} +4\Gamma (1+\tan^2{\phi})  \right)}{4 s (s+\Gamma) \tan^2{\phi}}&\;\;\;\;\;\;
\end{IEEEeqnarray}

These equations define a linear model for the amplitude and phase perturbations associated to a Duffing resonator in closed loop. Since this study focuses on the thermomechanical limits, we need to calculate the model inputs that account for thermomechanical noise. In the limit $k_B T \gg \hslash \omega_0$, where $\hslash$ is the reduced Planck constant, this is achieved by considering a random force with the PSD specified by \eqref{eq:S_thm}. When this random force is superimposed on a harmonic actuation force, the result is a quasi-harmonic total force with amplitude and phase random perturbations. We can obtain the PSD of the force phase perturbations $S_{\phi|\textrm{thm}}$ as the PSD of the thermomechanical force divided by the mean value of the force carrier squared \cite{rubiola2008phase}, $a_g^2 k_1^2/2$. Using \eqref{eq:fixed_point_equations_1}, we obtain
\begin{equation}\label{eq:Sph_thm}
    S_{\phi|\textrm{thm}}=\frac{2 S_{\textrm{thm}}}{a_g^2 k_1^2}= \frac{8 k_B T Q \sin^2 \phi}{m \omega_0^3 a_x^{2}}.
\end{equation}
Since thermomechanical noise is additive and uncorrelated with the force carrier, the PSD of the normalized amplitude perturbations will be the same as for the phase perturbations \cite{rubiola2008phase}. The non-normalized version to be used as input in our model results from multiplying $S_{\phi|\textrm{thm}}$ by the carrier amplitude squared, $a_g^2$, which gives
\begin{equation}\label{eq:Sa_thm}
    S_{a|\textrm{thm}}=\frac{2 S_{\textrm{thm}}}{k_1^2}=\frac{8 k_B T}{Q m \omega_0^3}.
\end{equation}

With the PSD of the inputs, we can calculate the PSD of the output amplitude and phase perturbations, $S_{ax}$ and $S_{\phi x}$ respectively, through the relation
\begin{equation}\label{eq:power_cl}
    \begin{bmatrix}
        S_{ax}(\omega)\\
        S_{\phi x}(\omega)
    \end{bmatrix}=
    \begin{bmatrix}
        |J_{11}(i\hat{\omega})|^2 & |J_{12}(i\hat{\omega})|^2 \\
        |J_{21}(i\hat{\omega})|^2 & |J_{22}(i\hat{\omega})|^2
    \end{bmatrix}
        \begin{bmatrix}
        S_{a|\textrm{thm}} \\
        S_{\phi|\textrm{thm}}
    \end{bmatrix},
\end{equation}
where $i$ is the imaginary unit.
With the last equation, the amplitude and phase noise of the resonator's displacement in a closed-loop configuration can be obtained. The latter is the relevant quantity for the present study as it determines the Allan deviation of the displacement signal, which is a measure of the minimum resonance frequency shift that can be resolved from the resonator in closed loop. With this model, an analysis of the displacement phase noise $S_{\phi x}$ as a function of the phase shift $\phi$ and the displacement amplitude $a_x$ is made in Supplementary Note \ref{note:allan}. The results indicate that the Allan deviation of a Duffing resonator can be approximated by three asymptotes at different integration time ranges: 
\begin{IEEEeqnarray}{RRLCL}
\label{eq:sigma_c_main}
    \sigma_y(&\tau&\ll\tau_c/\chi)&=& \sigma_{y0}\\
\label{eq:sigma_b_main}
    \sigma_y(\tau_c/\chi\ll&\tau&\ll\tau_c)&=&\sigma_{y0}\chi \frac{\pi}{\sqrt{3}}\frac{f_0\tau}{Q}\\
\label{eq:sigma_a_main}
    \sigma_y(&\tau&\gg\tau_c)&=& \sigma_{y0}\chi.
\end{IEEEeqnarray}
Each asymptote is expressed as a function of the Allan deviation for linear resonators actuated at resonance $\sigma_{y0}$ (\eqref{eq:sigma_linear}) and a factor $\chi$ that accounts for the amplitude-phase noise conversion. Defining the normalized amplitude $\hat{a}=a_x / a_{\rm{crit}}$, where $a_{\rm{crit}}=\sqrt[4]{64/27}/\sqrt{Q\gamma}$ is the critical amplitude \cite{febbo2013critical}, this factor can be expressed as
\begin{equation}
    \label{eq:chi}
    \chi=\sqrt{\frac{16}{3} \hat{a}^4   -\frac{8}{\sqrt{3}\tan{\phi}} \hat{a}^2 +\frac{1}{\sin^2{\phi}}}.
\end{equation}
The integration time that defines the boundaries between the different $\tau$ ranges is
\begin{equation}
    \label{eq:tau_c_main}
    \tau_c=\frac{\sqrt{3}}{\pi}\frac{Q}{f_0}.
\end{equation}

An important observation from this analysis is that the Allan deviation for $\tau\ll\tau_c/\chi$ depends only on the displacement amplitude $a_x$ and not on the particular choice of $\phi$ and $a_g$ to achieve that value of $a_x$. This implies that if the analysis is performed at fixed force amplitude $a_g$, the displacement amplitude $a_x$ and therefore also the Allan deviation at fast sensing depend on the resonator's phase shift $\phi$. Therefore, for the sake of simplicity and without loss of generality, the Allan deviation will only be analyzed at fixed displacement amplitude hereafter.

The analysis also shows that the smallest possible value of $\chi$ is $1$ and therefore the Allan deviation is larger than $\sigma_{y0}$ at $\tau\gg\tau_c$ if $\chi > 1$, see \eqref{eq:sigma_a_main}. However, $\phi$ can be chosen such that $\chi=1$ for any displacement amplitude (see Fig.~\ref{fig:chi}). From \eqref{eq:chi}, we can obtain an analytic expression for this optimal phase point  $\phi^*$:
\begin{equation}
    \label{eq:magic_angle}
    \phi^*=\arctan\left(\frac{\sqrt{3}}{4 \hat{a}^2} \right).
\end{equation}
If this phase shift is fixed by the feedback, the Allan deviation will be the same as for a linear resonator oscillating at resonance at the corresponding $a_x$, for any integration time. This condition has been previously referred to as the amplitude detachment point \cite{villanueva2013surpassing}. Several works have pointed out the possibility of tuning the phase of a Duffing resonator in order to reduce the phase noise \cite{villanueva2013surpassing,mestrom2009phase,kenig2012optimal, yurke1995theory}. These efforts discussed several cases that are particularly interesting for improving the stability of time references. Nevertheless, fixing this phase condition entails difficulties that might prevent its practicality in sensing. First, as visualized in Fig.~\ref{fig:chi}, $\chi$ becomes extremely sensitive to $\phi$ in the vicinity of $\phi=\phi^*$ as $\hat{a}$ becomes greater than 1. Second, for a given displacement setpoint $a_x$, a higher force is needed at phase $\phi=\phi^*$ than at $\phi=-\pi/2$. By using \eqref{eq:fixed_point_equations_1} and \eqref{eq:magic_angle}, it is deduced that the force needed to drive the resonator at $\phi^*$ is  higher than that needed at $\phi=-\pi/2$ by a factor that grows with $a_x^2$, see Fig.~\ref{fig:chi} (d). The higher force amplitude can introduce higher noise, which might prevent reaching the thermomehcanical limit. For these practical reasons, the focus of the present work is on Duffing resonators operated at resonance, i.e. $\phi=-\pi/2$. This condition minimizes the influence of small deviations of $\phi$ on sensor resolution, and maximizes the conversion of force to displacement.

Under the oscillation condition $\phi=-\pi/2$ and assuming a displacement amplitude well above the critical amplitude ($\hat{a}\gg \sqrt[4]{3}/2$), then $\chi \approx 4 \hat{a}^2/\sqrt{3}$. A graphic representation of \eqref{eq:sigma_c_main}, \eqref{eq:sigma_b_main} and \eqref{eq:sigma_a_main} under these assumptions is shown in Fig.~\ref{fig:model}(b). As seen, the Allan deviation reaches a minimum in the range $\tau\ll\tau_c$. This minimum value can be found by evaluating $\sigma_{y0}(\tau)$ at the boundary between the short and intermediate integration times, i.e. $\tau=\tau_c/\chi$:
\begin{equation}\label{eq:sigma_min}
    \sigma_{y\textrm{|min}}(\tau\ll\tau_c)=\frac{\sqrt[4]{3}}{2}\sqrt{ \frac{k_B T \gamma}{Q k_1}}.
\end{equation}
It is worth pointing out that the integration time at which this minimum Allan deviation is reached ($\tau_c/\chi$) is inversely proportional to the displacement amplitude at which the resonator is driven. On the contrary, the value of $\sigma_{y\textrm{|min}}(\tau\ll\tau_c)$ is a characteristic of the resonator, and does not depend on the oscillation amplitude set by the closed-loop controller. As a result, the model shows that the frequency resolution of a Duffing resonator presents a lower boundary at short gate times ($\tau\ll\tau_c$) given by the Duffing coefficient, the linear stiffness and the quality factor. Importantly, higher quality factors lead to resonators with better (lower) resolution limits at this $\tau$ regime, because at constant $k_1$, $T$, and $\gamma$, the value of the minimum Allan deviation for any $a_x$ is proportional to $Q^{-1/2}$. This conclusion extends on that from previous works \cite{roy2018improving,demir2019fundamental}, which were based on the assumption that a minimum Allan deviation curve was reached at a certain displacement amplitude, with that curve maintaining the $\sigma_y\propto \tau^{-1/2}$ form of linear resonators. In contrast, the analysis provided here shows that the amplitude-phase noise conversion, described by \eqref{eq:cl3}, depends not only on the driving amplitude but also on the frequency of the amplitude and phase noise, defined as an offset from the carrier frequency. As a consequence, the excess Allan deviation due to the amplitude-phase noise conversion depends on the integration time as well, yielding a minimum Allan deviation given by \eqref{eq:sigma_min} for $\tau\ll\tau_c$. The model also shows that, for long averaging times $\tau\gg\tau_c$, the minimum Allan deviation presents indeed a $\tau^{-1/2}$ dependence which is reached at a displacement amplitude close to the critical value, in accordance with previous works \cite{roy2018improving,demir2019fundamental}. Specifically, it is found that \eqref{eq:sigma_a_main} is minimized for $\hat{a}=\sqrt[4]{3}/2\approx 0.66$, which results in a minimum Allan deviation
\begin{equation}\label{eq:sigma_min_long_tau}
    \sigma_{y\textrm{|min}}(\tau\gg\tau_c) =\sqrt{\frac{3 k_B T \gamma}{k_1 \omega_0 \tau}}.
\end{equation}

To compare the theoretical model to the experimental results in Fig.~\ref{fig:results}(e), the Allan deviation limit calculated with \eqref{eq:sigma_min} is plotted as a horizontal dashed line. The experimental data for high actuation forces is in qualitative agreement with the model at $\tau\ll\tau_c$, with a diminishing Allan deviation at short integration times and an increasing Allan deviation at higher values of $\tau$. In between these regions a minimum value is found that does not depend on the driving force and shows good agreement with the value calculated with \eqref{eq:sigma_min}. For integration times longer than $\tau_c$, the $\tau^{-1/2}$ dependence is seen, consistent with \eqref{eq:sigma_min_long_tau}.
 

\begin{figure*}
\centering
\includegraphics[width=\textwidth]{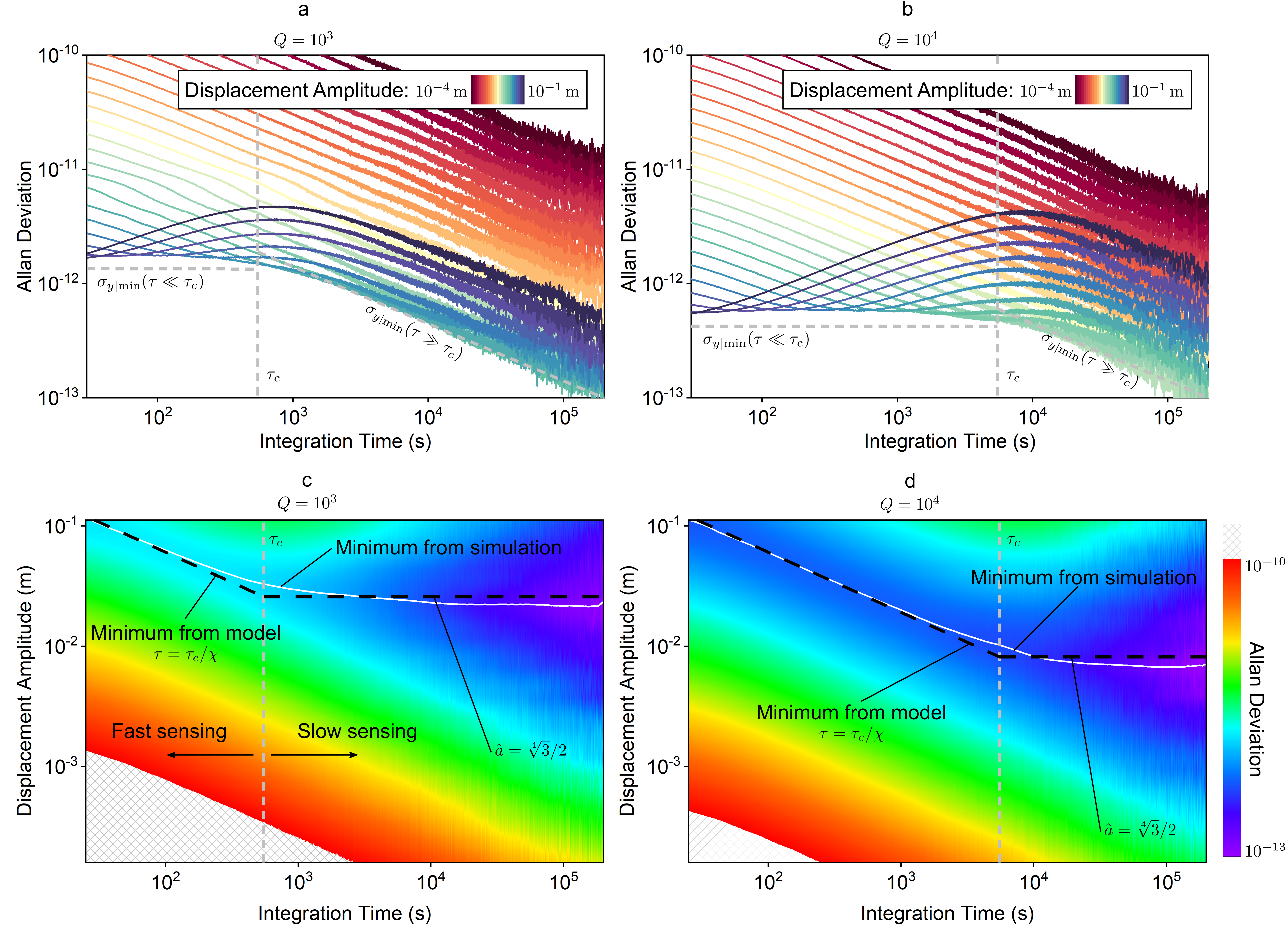}
\caption{Simulated Allan deviation for a Duffing resonator embedded in a direct feedback oscillator. The resonator parameters are $f_0=\SI{1}{\hertz}$, $k_1=\SI{1}{\newton\per\meter}$, $\gamma=\SI{1}{\per\square\meter}$. The temperature is assumed $\SI{300}{\kelvin}$. (a,b) Each curve represents the Allan deviation as a function of the integration time for a different displacement amplitude in the range \SI{e-4}{}--\SI{e-1}{\metre} for $\phi=-\pi/2$ and a quality factor of (a) \SI{e3}{} and (b) \SI{e4}{}. The minimum Allan deviation found by the model is indicated for each case by grey dashed lines. (c, d) Color maps based on the same simulations. The Allan deviation is represented as a function of the displacement amplitude and integration time for a quality factor of (c) \SI{e3}{} and (d) \SI{e4}{}. The white solid lines (smoothed) mark the displacement amplitude that minimizes the Allan deviation for each integration time simulated. The black dashed lines indicate the optimal displacement predicted by the model.}
\label{fig:simulations}
\end{figure*}

\section*{Simulations}

A Duffing resonator embedded in a direct feedback oscillator (DFO) was simulated in Matlab to validate the analytical model. The full non-linear dynamics of the resonator, as described by \eqref{eq:duffing_res}, was used in the simulations, with the details found in Methods. Fig.~\ref{fig:simulations} shows the simulation results for different oscillation amplitudes and $\phi=-\pi/2$, for $Q=\SI{e3}{}$ in Fig.~\ref{fig:simulations}(a,c) and $Q=\SI{e4}{}$ in Fig.~\ref{fig:simulations}(b,d). The other resonator parameters can be found in the figure caption. The results show good qualitative agreement with the model prediction for the minimum Allan deviation limits. The minimum Allan deviation for $\tau\ll\tau_c$ is confirmed as being independent of $\tau$, but dependent on the quality factor. In Figs.~\ref{fig:simulations}(c,d), a color map is shown to visualize the locus of the optimal displacement amplitudes that minimize the Allan deviation for each $\tau$, also in good agreement with the model. As seen, the optimal displacement amplitude is constant for $\tau\gg\tau_c$ but dependent on $\tau$ for $\tau\ll\tau_c$. As expected, the simulation results show that the model is not accurate in the vicinity of $\tau=\tau_c$. There, the Allan deviation slightly surpasses the limits defined in both regimes and the optimal displacement amplitude smoothly transitions between both regimes' predictions.

\section*{Discussion}

The obtained results provide new insights into the ultimate limits and optimum design of closed-loop resonant sensors that operate at their resonance frequency ($\phi=-\pi/2$). Particularly, in the fast-sensing regime, where the acquisition rate $f_s=1/\tau$ of the sensor needs to exceed $1/\tau_c$, sensor optimization needs to proceed along different lines, and operation in the non-linear regime becomes favourable. The oscillation amplitude that minimizes the Allan deviation in this regime for a given $f_s$ can be found by setting the acquisition rate to $f_s=\chi/\tau_c$. Using \eqref{eq:chi} for displacements amplitudes well above the critical amplitude ($\hat{a}\gg \sqrt[4]{3}/2$) and \eqref{eq:tau_c_main}, the optimal displacement normalized to the critical amplitude is found 
\begin{equation}\label{eq:ax_opt_2}
    \hat{a}_{\textrm{fast}}= \sqrt{\frac{3 Q f_s}{2 \omega_0}},
\end{equation}
with a non-normalized value of
\begin{equation}\label{eq:ax_opt}
    a_{x|\textrm{fast}}=\hat{a}_{\textrm{fast}}a_{\rm{crit}}=\frac{2}{\sqrt[4]{3}}  \sqrt{\frac{f_s}{\omega_0 \gamma}}.
\end{equation}
For this situation, the ultimate frequency resolution is given by \eqref{eq:sigma_min} and therefore can be minimized by increasing the quality factor and the linear stiffness,  and decreasing the cubic stiffness. Decreasing the acquisition rate does not further improve the minimum frequency resolution, unless it is increased sufficiently, where for $f_s = 1/\tau_c$ the slow sensing regime is reached.

In the slow-sensing regime, the displacement amplitude that yields the minimum Allan deviation can be obtained by minimizing \eqref{eq:sigma_a_main}, giving a normalized value of
\begin{equation}\label{eq:ax_opt_4}
    \hat{a}_{\textrm{slow}}= \frac{\sqrt[4]{3}}{2},
\end{equation}
and a non-normalized value of
\begin{equation}\label{eq:ax_opt_3}
    a_{x|\textrm{slow}}=\hat{a}_{\textrm{slow}} a_{\rm{crit}} =\sqrt{\frac{2}{3 Q \gamma}}.
\end{equation}

In this case, the ultimate frequency resolution is described by \eqref{eq:sigma_min_long_tau}, and therefore can be minimized by increasing the resonance frequency and the linear stiffness, and decreasing the cubic stiffness. Noticeably, increasing the quality factor does not further improve the frequency resolution, but lowering the acquisition rate, i.e. increasing the integration time, does help. 

Let us now reflect on the underlying mechanisms behind our findings. We have identified two main time-constants of importance. First we have the constant $\tau_c$, which is very close to the characteristic time the linear resonator needs to reach steady state, i.e. $Q/f_0$. Secondly, we have $\tau_c/\chi$, which characterizes the effect of increasing the displacement amplitude $a_x$. This time-constant results from a trade-off between a relative reduction of the input phase noise when increasing $a_x$, and the enhancement of phase noise by the amplitude-phase noise conversion due to the non-linear stiffness term $(3/8)\gamma a_x^2$ of \eqref{eq:slowdyn2}. The combination of both contributions has a minimum at $\tau=\tau_c/\chi$, that provides an optimal operation point for Duffing resonators at the fast sensing regime.

If a high acquisition frequency is not required, the oscillation amplitude $a_x$ can be reduced, increasing $\tau_c/\chi$ at constant minimum Allan deviation. This can continue until $a_x=a_{x|\rm{slow}}$ is reached, which presents the optimal situation for the slow sensing regime and results in $\chi=\sqrt{2}$. The remarkable fact is that if the acquisition frequency is decreased further ($f_s < 1/\tau_c$) at constant displacement amplitude $a_x$, it does not anymore result in an increase of the Allan deviation. Instead, the Allan deviation reduces proportionally to $f_s^{1/2}$ like in the linear regime. This behaviour at the slow sensing regime might be explained by the fact that the integration times are significantly longer than the characteristic time constant of the resonator, $Q/f_0$. All force fluctuations at shorter time scales are averaged out, with the resonator acting as a low-pass filter for the amplitude and phase perturbations.

For completeness, it must be noted that the closed-loop tracking system for the resonance frequency behaves as a low-pass filter for the phase noise, and can be described by additional time-constants \cite{olcum2015high}. These are set by the proportional-integral controller in a PLL, or by the amplifier bandwidth in a DFO. Nevertheless, a prerequisite for a closed-loop resonant sensor is that the resonance frequency tracking must be faster than the sensor acquisition rate, i.e. $f_s \ll 1/\tau_{cl}$, where $\tau_{cl}$ is the lowest time-constant of the close-loop phase-space transfer function of the controller. Therefore, for a properly designed closed-loop system, the controller dynamics can be disregarded when analyzing the resolution limits \cite{demir2019fundamental}.

The simulation results show, in accordance with the analytical model, that the minimum Allan deviation is reached for a different curve of Figs.~\ref{fig:simulations}(a,b), i.e.~different amplitude, at each integration time in the fast sensing regime. On the contrary, for slow sensing the minimum is reached for each integration time by the same curve, i.e.~same amplitude. Plotting the locus of these minima results in the asymptotic behaviour shown in Fig.~\ref{fig:intro}(c). It follows that increasing the quality factor of a Duffing resonator has two beneficial effects. First, the resonator will be ultimately more sensitive at the fast sensing regime. Second, the integration time defining the fast sensing regime will become longer, enabling better Allan deviations over a larger range of $\tau$. The ultimate frequency resolution at the slow sensing regime will be unaltered by the change in the quality factor.

Another important consideration is that the minimum Allan deviation limit found for fast sensing, $\sigma_{y\textrm{|min}}(\tau\ll\tau_c)$, can be reached for any integration time (acquisition rate) in a Duffing resonator, by tuning the displacement amplitude according to \eqref{eq:ax_opt}. This means that this frequency resolution can be achieved for arbitrarily high sensing speeds, assuming the controller can be made fast enough. The higher the required speed (shorter integration time), the higher the displacement amplitude needed. In real resonators, non-linear behaviours beyond the Duffing model, such as non-linear damping, will emerge for high enough displacements, which presumably will limit the sensing speed at which $\sigma_{y\textrm{|min}}(\tau\ll\tau_c)$ can be attained.

Finally, mass sensing is an application of utmost importance for the mechanical resonator community. In this context, the mass resolution $\delta_m$ can be related to the frequency resolution (Allan deviation) through the expression $\delta_{m}(\tau)=2 m \sigma_y(\tau)$ \cite{ekinci2004ultimate}. Then, the minimum mass resolution for fast sensing can be obtained by using \eqref{eq:sigma_min}:
\begin{equation}\label{eq:mass_sigma}
    \delta_{m\textrm{|min}}(\tau\ll\tau_c)=\frac{\sqrt[4]{3}}{\omega_0}\sqrt{\frac{k_B T m \gamma}{Q}}.
\end{equation}
Likewise, for slow sensing \eqref{eq:sigma_min_long_tau} gives
\begin{equation}\label{eq:mass_sigma_long}
    \delta_{m\textrm{|min}}(\tau\gg\tau_c)=\sqrt{\frac{12 k_B T m \gamma}{\omega_0^3 \tau}}.
\end{equation}

These two expressions can be applied to specific types of vibration modes, for which analytical expressions for $\gamma$ can be derived in order to optimize designs for best mass resolution. In Supplementary note \ref{sec:bending}, the specific case of doubly-clamped beams is analyzed.

\section*{Conclusion}
Previous knowledge suggested that the best frequency resolution of a resonant sensor with Duffing non-linearity operated at resonance is that attained when the oscillation amplitude is set at the onset of non-linearity. We present evidence that operation at higher amplitudes leads to lower frequency resolution at high acquisition rates $f_s \gg 1/\tau_c$. Whereas the precision of slow resonant sensors ($f_s \ll \tau_c$) can theoretically be improved indefinitely by decreasing the acquisition rate, the precision of fast resonant sensors ($f_s \gg \tau_c$) is fundamentally limited by a lower boundary independent of the acquisition rate, as a result of reaching the Duffing regime. In addition, we found that the frequency resolution limit at fast sensing is inversely proportional to the quality factor of the resonator. The reported experimental results, analytical model and numerical analysis provide a roadmap for device optimization and concept development to drive down the performance limits of resonant sensors. This understanding is expected to impact sensing applications requiring resolution and speed simultaneously, such as the mass characterization of bio-molecules with high throughput.

\section*{Methods}

\subsection{Device fabrication}
For the fabrication of \SI[product-units=power]{2 x 2}{\milli\metre\squared} nanomechanical membranes, a \SI{92}{\nano\meter}-thick \ce{Si3N4} film with \SI{1.1}{\giga\pascal} tensile stress was deposited by low pressure chemical vapor deposition (LPCVD) on a silicon wafer. After this, the wafer was diced into \SI[product-units=power]{10 x 10}{\milli\metre\squared} chips. Membrane patterns were defined on one of the chips with ebeam lithography on a positive tone resist (AR-P 6200). The \ce{Si3N4} film was then patterned by inductively coupled plasma reactive ion etching (ICP-RIE) based on \ce{CHF3}. The resist was then removed by dimethylformamide, next the organic residues were cleaned with piranha, and the remaining surface oxides were removed with hydrofluoric acid solution. To facilitate the release of the membrane, a square array of circular holes with radius \SI{0.75}{\micro\meter} and center-to-center distance \SI{3.5}{\micro\meter} in both x and y directions was patterned. As the last step, the \ce{Si3N4} membranes were released from the Si substrate using ICP isotropic etching with \ce{SF6} at \SI{-120}{\degree}.
A circular-shaped pedestal, located in the middle of the membrane, was left in the Si substrate as a position indicator, with a top-surface radius of \SI{10}{\micro\meter} and disconnected from the suspended membrane.

\subsection{Experimental setup}

The silicon chip containing the membrane was placed on a piezo-actuator for the experiments. All the measurements were performed inside a vacuum chamber at a pressure of \SI{2}{\milli\pascal}. Thanks to a glass window on the vacuum chamber, the vibrations were recorded by a Doppler laser vibrometer (Polytec MSA400). The actuation signal was provided to the piezo-actuator by a digital lock-in (Zurich Instruments HF2LI). The same instrument was used to record the displacment signal from the vibrometer. The PSD of the displacement was first recorded with no signal applied to the piezo-actuator (Fig.~\ref{fig:results}(c)). Then, frequency sweeps around the resonance were performed for various voltage amplitudes applied to the piezo-actuator (Fig.~\ref{fig:results}(d)). The resonator's transfer function, obtained through fitting the data from these experiments, allowed to obtain the conversion factor from voltage to force associated to the piezo-actuator-chip assembly. For the Allan deviation experiments, the digital lock-in was configured in PLL mode, with a phase detector filter of order 8 and bandwidth \SI{18}{\kilo\hertz}. The proportional-integral controller was set with a proportional constant of \SI{1.8e3}{\hertz} and an integral constant of \SI{5.4e3}{\hertz\squared}, ensuring enough PLL bandwidth to characterize integration times longer than \SI{100}{\micro\second}. The target phase was set to the phase shift featured by the resonator at the frequency of maximum amplitude when actuated in the linear range, which corresponds to the theoretical point $\phi=-\pi/2$. Under these conditions, data series of the resonator phase were recorded for 5 minutes with an acquisition rate of \SI{7.2}{\kilo\hertz}, while applying the force amplitudes shown in Fig.~\ref{fig:results}(d). White noise with bandwidth of \SI{9}{\mega\hertz}, provided by a waveform generator (Agilent 33220A), was introduced in the digital lock-in and superimposed to the actuation signal. For each force amplitude, the Allan deviation was calculated from the phase data in Matlab following the procedure in \cite{manzaneque2020method}.

\subsection{Simulations}

A DFO was simulated in Matlab/Simulink including the full non-linear differential equation of a Duffing resonator. The DFO was formed by the resonator and an amplifier-phase shifter chain, which generates a force proportional to the resonator's displacement. The amplifier gain was set such that the closed-loop gain of the system is slightly above 1. From \eqref{eq:fixed_point_equations_1}, this is achieved by an amplifier gain of $-k_1 /(Q \sin{\phi})$. For stable oscillations, a saturation element was included in the feedback to set the oscillation amplitude to an arbitrary value $a_x$. The phase shifter was fixed to $\pi/2$, which ensures the Barkhausen criterion is met for a resonator phase $\phi=-\pi/2$ and the oscillation frequency is $\omega_p$. The amplifier, phase shifter and saturation were modelled as noiseless elements with infinite bandwidth. To include the thermomechanical noise affecting the resonator, a random force with white PSD in accordance to \eqref{eq:S_thm} was superimposed to the force applied by the feedback to the resonator. The phase of the resonator displacement was calculated through a phase detector scheme and used to obtain the Allan deviation following the procedure in \cite{manzaneque2020method}. The simulated time was $\SI{2e5}{\second}$ and the time step was fixed to the inverse of the linear resonance frequency divided by 100. The default fixed-step solver of Simulink was used.



\begin{acknowledgements}
This research work has received funding from the European Union’s Horizon 2020 research and innovation programme under the Marie Skłodowska-Curie grant agreement No 707404, and under Grant Agreement No. 881603 Graphene Flagship. FA acknowledges financial support from European Research Council (ERC) Starting Grant Number 802093. The opinions expressed in this document reflect only the authors' view. The European Commission is not responsible for any use that may be made of the information it contains.
This publication is also part of the project Probing the physics of exotic superconductors with microchip Casimir experiments (740.018.020) of the research programme NWO Start-up which is partly financed by the Dutch Research Council (NWO).  
\end{acknowledgements}

\section*{Bibliography}
\bibliographystyle{unsrt}
\bibliography{zHenriquesLab-Mendeley}

\onecolumn
\newpage


\captionsetup*{format=largeformat}


\counterwithin{figure}{section}
\renewcommand{\thefigure}{S\arabic{figure}}

\section{Linearized Model of the Amplitude-Phase Space of a Duffing Resonator in Closed Loop}
\label{note:linearization} 

The system of equations (\ref{eq:slowdyn1}-\ref{eq:slowdyn3}) can be linearized around a fixed point by a pair of values $a_x=a_{x0}$ and $\phi=\phi_0$, which unambiguously defines $a_g=a_{g0}$ and $\psi=\psi_0$ according to equations \eqref{eq:fixed_point_equations_1} and \eqref{eq:fixed_point_equations_2}. To express the linearized system in the state-space representation, the states are defined as the perturbations of the amplitude and phase of the resonator's displacement, $\Delta a_x$ and $\Delta \phi_x$, that coincide with the system outputs. The system inputs are identified as the perturbations of the amplitude and phase of the force acting on the resonator,  $\Delta a_g$ and $\Delta \phi_g$ respectively. This results in a system defined by the matrices $D$ and $E$:

\begin{equation}\label{eq:ss1}
    \begin{bmatrix} \dot{\Delta a_x} \\ \dot{\Delta \phi_x} \end{bmatrix}=
    \begin{bmatrix} D_{11}&D_{12} \\ D_{21}&D_{22} \end{bmatrix}
    \begin{bmatrix} \Delta a_x \\ \Delta \phi_x \end{bmatrix}
    +\begin{bmatrix} E_{11}&E_{12} \\ E_{21}&E_{22} \end{bmatrix}
    \begin{bmatrix} \Delta a_g \\ \Delta \phi_g \end{bmatrix},
\end{equation}
with the matrices elements given by
\begin{IEEEeqnarray}{RCL}
\label{eq:D11}
    D_{11}=& \left. \frac{\partial \eta_a}{\partial a_x}\right|_{a_{x0},\phi_0,a_{g0},\psi_0}&= -\Gamma
\\
\label{eq:D12}
    D_{12}=& \left. \frac{\partial \eta_a}{\partial \phi_x}\right|_{a_{x0},\phi_0,a_{g0},\psi_0}&= \frac{\Gamma a_{x0}}{\tan{\phi_0}}
\\
\label{eq:D21}
    D_{21}=& \left. \frac{\partial \eta_\phi}{\partial a_x}\right|_{a_{x0},\phi_0,a_{g0},\psi_0}&= \frac{3 }{4}\gamma a_{x0} -\frac{\Gamma}{a_{x0}\tan{\phi_0}}  
\\
\label{eq:D22}
    D_{22}=& \left. \frac{\partial \eta_\phi}{\partial \phi_x}\right|_{a_{x0},\phi_0,a_{g0},\psi_0}&= -\Gamma
\\
\label{eq:E11}
    E_{11}=& \left. \frac{\partial \eta_a}{\partial a_g}\right|_{a_{x0},\phi_0,a_{g0},\psi_0}&= -\frac{\sin{\phi_0}}{2}
\\
\label{eq:E12}
    E_{12}=& \left. \frac{\partial \eta_a}{\partial \phi_g}\right|_{a_{x0},\phi_0,a_{g0},\psi_0}&= -\frac{\Gamma a_{x0}}{\tan{\phi_0}}
\\
\label{eq:E21}
    E_{21}=& \left. \frac{\partial \eta_\phi}{\partial a_g}\right|_{a_{x0},\phi_0,a_{g0},\psi_0}&= -\frac{\cos{\phi_0}}{2 a_{x0}}
\\
\label{eq:E22}
    E_{22}=& \left. \frac{\partial \eta_\phi}{\partial \phi_g}\right|_{a_{x0},\phi_0,a_{g0},\psi_0}&= \Gamma.
\end{IEEEeqnarray}
This system can be represented in the Laplace domain to obtain
\begin{equation}\label{eq:laplace_sup}
    \begin{bmatrix} A_x(s) \\ \Phi_x(s) \end{bmatrix}=
    \begin{bmatrix} H_{11}(s)&H_{12}(s) \\ H_{21}(s)&H_{22}(s) \end{bmatrix}
    \begin{bmatrix} A_g(s) \\ \Phi_g(s) \end{bmatrix},
\end{equation}
where $A_x(s)$, $\Phi_x(s)$, $A_g(s)$ and $\Phi_g(s)$ are the Laplace transforms of $\Delta a_x$, $\Delta \phi_x$, $\Delta a_g$ and $\Delta \phi_g$ respectively. The transfer function matrix can be found by making $H(s)=\left( s \mathbb{I} - D \right)^{-1} E $, where $\mathbb{I}$ is the identity matrix of size 2 \cite{friedland2005control}. This yields
\begin{IEEEeqnarray}{CCL}
\label{eq:H11}
    H_{11}(s)&=&\frac{  s (\cos (2 \phi_0 ) -1) -2 \Gamma}{P\sin{\phi_0}}
    \\
\label{eq:H12}
    H_{12}(s)&=&-\frac{4 s \Gamma   a_{x0} }{P \tan{\phi_0}}
    \\
\label{eq:H21}
    H_{21}(s)&=&-\frac{4 s \cos \phi_0 +3 \gamma a_{x0}^{2} \sin{\phi_0} }{2P a_{x0} }
    \\
\label{eq:H22}
    H_{22}(s)&=&\frac{\Gamma  \left(4 \tan^2{\phi_0} (s+\Gamma)  -3 \gamma a_{x0}^{2} \tan{\phi_0} +4 \Gamma\right)}{P \tan^2{\phi_0}},
\end{IEEEeqnarray}
with $P= 4 (\Gamma +s)^2 + 4 \Gamma ^2 / \tan^2{\phi_0 }-3 \Gamma  \gamma a_{x0}^{2} / \tan {\phi_0 }$.

As discussed in the main text, a closed-loop control is needed to keep the resonator at the fixed point. Its effect on the amplitude and phase perturbations is accounted by a matrix $B(s)$ in the feedback path, as depicted in Fig.~\ref{fig:model}(a). For a phase-locked loop (PLL) scheme as used in the experiments, the phase perturbations are fed back by the phase detector-controller chain \cite{ manzaneque2020method}. For frequencies lower than the bandwidth of the PLL transfer function, the dynamic behaviour of the feedback path can be neglected \cite{demir2019fundamental}. This implies that the phase perturbations are passed from the input to the output of the controller with a gain of 1, as long as the PLL is faster than the maximum sensing speed of interest. On the other hand, the amplitude perturbations are rejected by the feedback. This results in a feedback matrix
\begin{equation}\label{eq:B_matrix}
    B(s) =
    \begin{bmatrix} 0 & 0 \\  0 & 1 \end{bmatrix}.
\end{equation}

Alternatively to the PLL scheme, direct feedback oscillators (DFO) also provide a means to keep a resonator oscillating at a desired fixed point \cite{villanueva2013surpassing}. This type of system results from using an amplifier and phase shifter to feedback the resonator's displacement into its force. The phase shift $\phi_0$ can be controlled by offsetting the loop phase through the phase shifter. Similarly to the PLL case, the feedback path of a DFO lets the phase perturbations pass with gain 1, for perturbation frequencies lower than the bandwidth of the amplifier. To control the amplitude of oscillation $a_{x0}$, either a non-linear saturation block or an automatic gain control (AGC) can be placed in the feedback path. The $B$ matrix of a DFO can be approximated by \eqref{eq:B_matrix} as long as the amplitude control method used does not let the amplitude perturbations pass. Therefore, an AGC must allow for slow variations of the amplitude to allow the startup of the oscillations, while stopping fast variations. That is to say, it must behave as a low-pass filter for the amplitude perturbations. For perturbation with frequencies higher than the bandwidth of the AGC, the feedback effectively stops the amplitude perturbations, and \eqref{eq:B_matrix} can be regarded valid. To summarize, the present analysis is also valid for DFO as long as the relevant frequencies of the amplitude and phase noise are higher than the bandwidth of the AGC but lower than the bandwidth of the amplifier. In the context of sensing, this implies that the AGC must be slower that the maximum integration time (minimum sensing speed) of interest, while the amplifier must be faster than the minimum integration time (maximum sensing speed) of interest.

Given the $H(s)$ and $B(s)$ matrices, the closed-loop transfer function matrix $J(s)$ that governs the dynamics of the amplitude and phase perturbations can be obtained by using the feedback equation
\begin{equation}\label{eq:feedback}
    J(s)=[I-H(s)B(s)]^{-1}H(s),
\end{equation}
which results in
\begin{IEEEeqnarray}{CCL}
\label{eq:cl1_sup}
    J_{11}(s)&=&-\frac{\sin \phi_0}{2 (s+\Gamma)}
    \\
\label{eq:cl2_sup}
    J_{12}(s)&=&-\frac{\Gamma a_{x0} }{\tan\phi_0(s+\Gamma)}
    \\
\label{eq:cl3_sup}
    J_{21}(s)&=&-\frac{3 \gamma a_{x0}^{2} \sin \phi_0 +4 s \cos \phi_0}{8 a_{x0} s ( s +  \Gamma)}
    \\
\label{eq:cl4_sup}
    J_{22}(s)&=&\frac{\Gamma \left(4 s \tan^2{\phi_0} -3 \gamma a_{x0}^{2} \tan{\phi_0} +4\Gamma (1+\tan^2{\phi_0})  \right)}{4 s (s+\Gamma) \tan^2{\phi_0}}.
\end{IEEEeqnarray}
For simplicity, $a_{x0}$ and $\phi_0$ have been renamed as $a_x$ and $\phi$ respectively in the main text as well as in the supplementary notes hereafter.

\section{Phase Noise and Allan Deviation of a Duffing Resonator in Closed Loop}
\label{note:allan}
The PSD of the displacement phase noise for a closed-loop Duffing resonator dominated by thermomechanical noise is obtained from \eqref{eq:power_cl}:
\begin{equation}\label{eq:phase_noise_1}
    S_{\phi x} (\omega) =|J_{21}(i\hat{\omega})|^2 S_{a|\textrm{thm}} +|J_{22}(i\hat{\omega})|^2 S_{\phi |\textrm{thm}}.
\end{equation}
The result can be expressed as
\begin{equation}\label{eq:phase_noise_2}
    S_{\phi x} (f) = \frac{G}{f^2}\cdot\frac{ f^2+\chi^2f_c^2} {f^2+f_c^2},
\end{equation}
where $f$ represents frequency offset from the carrier frequency, and the constants are given by
\begin{IEEEeqnarray}{RCL}
    \label{eq:phase_noise_3}
    G&=&\frac{ k_B T}{2 \pi^2 m Q  \omega_0 a_{x}^2}
    \\
    \label{eq:phase_noise_4}
    f_c&=& \frac{f_0}{2 Q}
    \\
    \label{eq:phase_noise_5}
    \chi&=&\frac{1}{2}\sqrt{9 \gamma^2 Q^2 a_x^4  +4 /\sin^2{\phi} -12\gamma Q a_x^2 / \tan{\phi}}.
\end{IEEEeqnarray}
The factor $\chi$ accounts for the amplitude-phase noise conversion arising from the Duffing characteristic or from operating the resonator at $\phi\neq -\pi/2$. A linear resonator ($\gamma=0$) operated at $\phi=-\pi/2$ gives $\chi=1$, which results in $ S_{\phi x} (f)=G/f^2$. Note that $\chi$ is real and larger than 1 for $\phi \in (-\pi,0)$. 

Assuming strong noise conversion ($\chi\gg 1$), three frequency ranges can be defined in which the derived phase noise is approximated by the asymptotes of \eqref{eq:phase_noise_2}:
\begin{IEEEeqnarray}{LCCC}
    \label{eq:phase_noise_6}
    S_{\phi x} (f\ll f_c)&=&G\chi^2&\frac{1}{f^2}
    \\
    \label{eq:phase_noise_7}
    S_{\phi x} (f_c\ll f \ll\chi f_c)&=&G  \chi^2 f_c^2 &\frac{1}{f^4}    \\
    \label{eq:phase_noise_8}
    S_{\phi x} (f\gg\chi f_c)&=&G&\frac{1}{f^2}.
\end{IEEEeqnarray}
Note that for $f\gg \chi f_c$ there is no effect of the amplitude-phase noise conversion, and the phase noise is the same as for a linear resonator operated at resonance.

The Allan deviation $\sigma_y$, can be obtained from the PSD of the displacement phase noise through the integral expression \cite{barnes1971characterization}
\begin{equation}\label{eq:ad_phase_noise}
    \sigma_{y}^2 (\tau)=2 \int_{0}^{+\infty} S_{\phi x}(f)  \frac{\sin^4\left(\pi f \tau \right)}{\left(  \pi f \tau \right)^2}  \text{d}{f},
\end{equation}
where $\tau$ is the integration time. Simplified expressions for its evaluation at each of the frequency ranges defined above can be found in \cite{barnes1971characterization}. The phase noise at each frequency range can be converted to Allan deviation for a corresponding integration time range. The phase noise for high frequencies ($f\gg\chi f_c$) determines the Allan deviation for short integration times below certain value $\tau_c /\chi$. Therefore, by introducing \eqref{eq:phase_noise_8} in \eqref{eq:ad_phase_noise}, we obtain
\begin{equation}\label{eq:sigma_c}
    \sigma_y(\tau\ll\tau_c/\chi)=\frac{1}{a_x} \sqrt{\frac{k_B T}{ m Q \omega_0^3  \tau }} = \sigma_{y0}.
\end{equation}
$\sigma_{y0}$ is the Allan deviation of a linear resonator operated at $\phi=-\pi/2$. Likewise, the phase noise for $f\ll f_c$ dominates the Allan deviation for integration times longer than $\tau_c$. Therefore, by introducing \eqref{eq:phase_noise_6} in \eqref{eq:ad_phase_noise}, we obtain
\begin{equation}\label{eq:sigma_a}
    \sigma_y(\tau\gg\tau_c)= \chi\sigma_{y0}.
\end{equation}
From the phase noise at intermediate frequencies, the Allan deviation at intermediate integration times is obtained. Therefore, by introducing \eqref{eq:phase_noise_7} in \eqref{eq:ad_phase_noise}, we obtain
\begin{equation}\label{eq:sigma_b}
    \sigma_y(\tau_c/\chi\ll\tau\ll\tau_c)= \frac{\pi}{\sqrt{3}}\frac{f_0\tau}{Q}\chi\sigma_{y0}.
\end{equation}
By making $\sigma_y(\tau\gg\tau_c)=\sigma_y(\tau_c/\chi\ll\tau\ll\tau_c)$ and resolving for $\tau$, it is found that
\begin{equation}\label{eq:tau_c}
    \tau_c=\frac{\sqrt{3}}{\pi}\frac{Q}{f_0}.
\end{equation}

\section{Resolution Limits of a Doubly-Clamped Beam}
\label{sec:bending}

We can particularize the mass resolution limits found to the widely studied resonators based on doubly-clamped beams. For a rectangular beam of length $l$, width $w$, thickness $h$, mass density $\rho$ and Young's modulus $Y$, we can express the Duffing coefficient associated to the fundamental mode as \cite{westra2012nonlinear}
\begin{equation}\label{eq:2c_beam_gamma}
    \gamma=\frac{2}{h^2},
\end{equation}
the linear stiffness as \cite{hauer2013general}
\begin{equation}\label{eq:2c_beam_k1}
    k_1=\frac{16 Y w h^3}{l^3},
\end{equation}
and the fundamental resonance frequency as \cite{blevins2001formulas}
\begin{equation}\label{eq:wn}
    \omega_0=\frac{\lambda_0^2 h}{l^2} \sqrt{\frac{Y}{12 \rho}} ,
\end{equation}
with $\lambda_0=4.73$. By combining these expressions with \eqref{eq:sigma_min}, we reach
\begin{equation}\label{eq:sigma_min_bridge}
     \sigma_{y\textrm{|min}}(\tau\ll\tau_c)=\frac{\sqrt[4]{3}}{\sqrt{32}}\sqrt{ \frac{k_B T l^3}{Q Y h^5 W}}.
\end{equation}
Doing the same with \eqref{eq:sigma_min_long_tau}, we reach
\begin{equation}\label{eq:sigma_min_long_tau_bridge}
     \sigma_{y\textrm{|min}}(\tau\gg\tau_c)=\frac{1}{2\lambda_0}  \sqrt[4]{\frac{27\rho}{Y^3}}\sqrt{ \frac{k_B T l^5}{h^6 w \tau}}.
\end{equation}
We can also calculate the minimum mass resolution at both regimes by making $\delta_m=2m \sigma_y$, with $m=k_1/\omega_0^2$:
\begin{equation}\label{eq:dm_min_bridge}
     \delta_{m\textrm{|min}}(\tau\ll\tau_c)=\frac{192\sqrt[4]{3}\rho}{\sqrt{8}\lambda_0^4}\sqrt{ \frac{k_B T l^5 w}{Q Y h^3}},
\end{equation}
\begin{equation}\label{eq:dm_min_long_taubridge}
     \delta_{m\textrm{|min}}(\tau\gg\tau_c)=\frac{192\sqrt[4]{27}}{\lambda_0^5} \sqrt[4]{\frac{\rho^5}{Y^3}}\frac{1}{h^2}\sqrt{ \frac{k_B T l^7 w}{\tau}}.
\end{equation}
These expressions establish the ultimate mass resolution of a doubly-clamped beam due its geometrical non-linearity, as a function of the material properties, dimensions, quality factor and temperature.

\section*{Supplementary Figures}

\newpage

\begin{figure*}
\centering
\includegraphics[width=\textwidth]{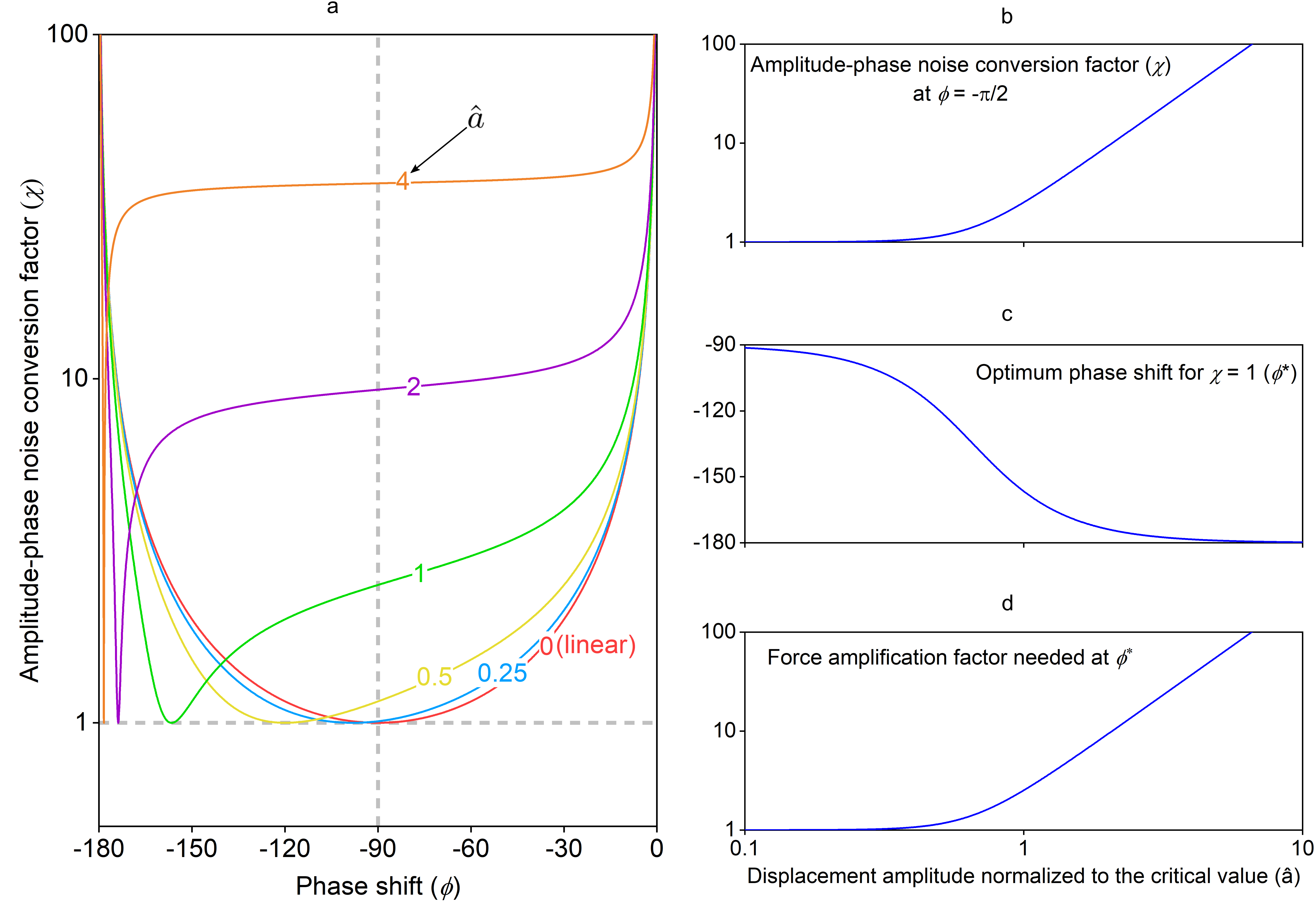}
\caption{(a) Factor $\chi$ accounting for the excess Allan deviation at $\tau\gg\tau_c$ due to the amplitude-phase noise conversion, with the phase shift $\phi$ in the horizontal axis. Each curve, obtained from \eqref{eq:chi}, represents a different value of the displacement amplitude normalized to the critical value ($\hat{a}$), indicated by labels. (b) Factor $\chi$ for a resonator operated at $\phi=-\pi/2$, with $\hat{a}$ in the horizontal axis. For low values of $\hat{a}$, a Duffing resonator behaves as a linear resonator, with $\chi\approx1$. (c) Optimum phase shift $\phi^*$ that minimizes $\chi$, with $\hat{a}$ in the horizontal axis. If $\phi=\phi^*$ is set, a Duffing resonator shows the same Allan deviation as a linear resonator operated at resonance for a given displacement amplitude $a_x$. (d) Force needed to set a given value of $\hat{a}$ with $\phi=\phi^*$, divided by the force needed to to set the same value of $\hat{a}$ with $\phi=-\pi/2$. This curve is obtained from \eqref{eq:fixed_point_equations_1}}.
\label{fig:chi}
\end{figure*}

\end{document}